%% file: main.tex
\documentclass[a4paper,reqno,twoside]{amsart}
\usepackage{bbm}
\usepackage[left=3.5cm,right=3.5cm,top=3.5cm,bottom=3.5cm,footskip=1.5cm,headsep=1cm,bindingoffset=0.cm,]{geometry}

\input{mainsetup}

\input{specifics}

\input{commands}

\newcommandx{\silvio}[2][1=]{\todo[linecolor=blue,backgroundcolor=blue!25,bordercolor=blue,#1]{Silvio: #2}}

\newcommandx{\alex}[2][1=]{\todo[linecolor=red,backgroundcolor=red!25,bordercolor=red,#1]{Alex: #2}}

\title[A Real-Space Formulation of the Zak Phase via Weyl $m$-Functions]{A Real-Space Formulation of the Zak Phase via Weyl $m$-Functions}

\begin{document}
\author[H. Ammari]{Habib Ammari \,\orcidlink{0000-0001-7278-4877}}
\address{\parbox{\linewidth}{Habib Ammari\\
 ETH Z\"urich, Department of Mathematics, Rämistrasse 101, 8092 Z\"urich, Switzerland, \href{http://orcid.org/0000-0001-7278-4877}{orcid.org/0000-0001-7278-4877}}.}
 \email{habib.ammari@math.ethz.ch}
 \thanks{}

\author[C. Thalhammer]{Clemens Thalhammer}
\address{\parbox{\linewidth}{Clemens Thalhammer\\
 ETH Z\"urich, Department of Mathematics, Rämistrasse 101, 8092 Z\"urich, Switzerland.}}
 \email{clemens.thalhammer@sam.math.ethz.ch}

 \begin{abstract}
We establish a new, real-space formula for the Zak phase for one dimensional periodic Jacobi operators in terms of the Weyl $m_+$-function that does not rely on Floquet-Bloch theory. This novel representation highlights the dependence of the Zak phase on boundary terms. Moreover, we show how to recover the classical quantisation of the Zak phase for periodic Jacobi operators with inversion symmetric fundamental cells. 
 \end{abstract}

\maketitle


\bigskip

\\

\tableofcontents

\input{introduction}
\input{preliminaries}
\input{setup_and_main_result}

\input{numerics}
\input{proof_main_result}
\input{conclusion}

\section{Acknowledgements}
This work was supported in part by the Swiss National Science Foundation grant number 200021-236472. The authors thank  Jiayu Qiu and Alexander Uhlmann for many useful discussions.

\section{Code Availability}
The code used to generate figures \ref{fig: RiceMeleComparison}, \ref{fig: trimer} is openly available at \url{https://github.com/cthalhammer/ZakPhaseWeyl}.

\appendix

\printbibliography

\end{document}

%% file: mainsetup.tex
\usepackage[utf8]{inputenc}
\usepackage[english]{babel}
\usepackage[T1]{fontenc} % To switch to the T1 encoding
\usepackage{lmodern} % To switch to Latin Modern
\rmfamily % To load Latin Modern Roman and enable the following NFSS declarations.
\DeclareFontShape{T1}{lmr}{b}{sc}{<->ssub*cmr/bx/sc}{}
\DeclareFontShape{T1}{lmr}{bx}{sc}{<->ssub*cmr/bx/sc}{}

\numberwithin{equation}{section}
\usepackage{amsmath}
\usepackage{amssymb}
\usepackage{amsthm}
\usepackage{amsfonts}
\usepackage{mathtools}

\usepackage{mathdots}
\usepackage{caption}
\usepackage{subcaption}

\usepackage{dsfont} % for 1 bbmath
\usepackage[format=hang]{caption}
\usepackage[normalem]{ulem}

\usepackage{standalone}
\usepackage{graphicx}
\usepackage{imakeidx}
\makeindex
\usepackage{setspace}
\usepackage{appendix}
\usepackage{pdfpages}
\usepackage{upgreek}
\usepackage{tabulary}
\usepackage{booktabs}
\usepackage{extarrows}
\usepackage{tabularx}
\usepackage{float}
\restylefloat{table}
\usepackage{enumitem}
\usepackage{csquotes}
\usepackage{bm}
\usepackage{orcidlink}
\usepackage{lipsum}

\usepackage{tikz}
\usepackage{tikz-layers}
\usepackage{tikz-cd}
\usepackage[arrow, matrix, curve]{xy}
\usetikzlibrary{matrix,positioning,decorations.pathreplacing,calc}
\usetikzlibrary{
    decorations.text,%
    decorations.markings,%
    shadows}
\usetikzlibrary{calc,intersections}
\usepackage{adjustbox}
\usepackage{graphicx}

\usepackage[
    style=numeric,
    sorting=nty,
    maxnames=99,
    maxalphanames=5,
    natbib=true,
    backend=biber,
    sortcites]{biblatex}

\DeclareNameAlias{default}{family-given}

\AtEveryBibitem{% Clean up the bibtex rather than editing it
    \clearfield{url}
    \clearfield{issn}
    \clearfield{isbn}
    \clearfield{urldate}

    \ifentrytype{book}{
        \clearfield{pages}}{% 
    }
}

\usepackage{xargs}                      % Use more than one optional parameter in a new commands
\usepackage[prependcaption,textsize=tiny,textwidth=3cm]{todonotes}
\newcommandx{\unsure}[2][1=]{\todo[linecolor=red,backgroundcolor=red!25,bordercolor=red,#1]{#2}}
\newcommandx{\change}[2][1=]{\todo[linecolor=blue,backgroundcolor=blue!25,bordercolor=blue,#1]{#2}}
\newcommandx{\info}[2][1=]{\todo[linecolor=OliveGreen,backgroundcolor=OliveGreen!25,bordercolor=OliveGreen,#1]{#2}}
\newcommandx{\improvement}[2][1=]{\todo[linecolor=black,backgroundcolor=black!25,bordercolor=black,#1]{#2}}
\newcommandx{\thiswillnotshow}[2][1=]{\todo[disable,#1]{#2}}

\usepackage[noabbrev,capitalize]{cleveref}
\crefname{proposition}{Proposition}{Propositions}
\crefname{equation}{}{}

\newtheorem{theorem}{Theorem}[section]
\newtheorem{lemma}[theorem]{Lemma}

\theoremstyle{definition}
\newtheorem{definition}[theorem]{Definition}

\newtheorem{remark}[theorem]{Remark}

\crefname{assumption}{Assumption}{Assumptions}
\crefname{definition}{Definition}{Definitions}
\crefname{corollary}{Corollary}{Corollaries}
\crefname{enumi}{item}{items}

\creflabelformat{subfigure}{#2\textsc{#1}#3}

\usepackage[final]{microtype}

\makeatletter
\newsavebox\myboxA
\newsavebox\myboxB
\newlength\mylenA

\newcommand*\xoverline[2][0.75]{%
  \sbox{\myboxA}{$\m@th#2$}%
  \setbox\myboxB\null% Phantom box
  \ht\myboxB=\ht\myboxA%
  \dp\myboxB=\dp\myboxA%
  \wd\myboxB=#1\wd\myboxA% Scale phantom
  \sbox\myboxB{$\m@th\overline{\copy\myboxB}$}%  Overlined phantom
  \setlength\mylenA{\the\wd\myboxA}%   calc width diff
  \addtolength\mylenA{-\the\wd\myboxB}%
  \ifdim\wd\myboxB<\wd\myboxA%
    \rlap{\hskip 0.5\mylenA\usebox\myboxB}{\usebox\myboxA}%
  \else
    \hskip -0.5\mylenA\rlap{\usebox\myboxA}{\hskip 0.5\mylenA\usebox\myboxB}%
  \fi}
\makeatother

\usepackage{fancyhdr}

\fancypagestyle{plain}{%
    \fancyhead[C]{}
    \fancyfoot[C]{}
    
    }
\fancypagestyle{nosection}{%
    \fancyhead[CE]{}
    \fancyhead[CO]{}
    \fancyfoot[CE,CO]{\thepage}
}

%% file: specifics.tex
\usepackage{tikz}

\usetikzlibrary{matrix} 
\usetikzlibrary{arrows,arrows.meta,bending} %new code

\usepackage{nicematrix}

\usetikzlibrary{hobby}

\usepackage{calc}

\newcommand{\bseq}[2]{{\bm{#1}^{(#2)}}}

%% file: commands.tex
\renewcommand{\tilde}{\widetilde}

\renewcommand{\bar}[1]{\overline{#1}}

\renewcommand{\qedsymbol}{$\blacksquare$}

\newcommand{\nm}{\noalign{\smallskip}}
\newcommand{\ds}{\displaystyle}

\renewcommand{\epsilon}{\varepsilon}

\renewcommand{\tilde}{\widetilde}

\usepackage{tcolorbox}
\usetikzlibrary{tikzmark,calc}

%% file: introduction.tex
\section{Introduction}

Since the foundational work of Zak \cite{zak1989berry}, the Zak phase has emerged as a central concept in the study of topological properties of one-dimensional quantum systems. It captures the geometric phase accumulated by a Bloch wave function across the Brillouin zone and provides a direct link between the bulk properties of a material and its edge phenomena, a principle often referred to as the bulk-boundary correspondence. While the Zak phase can, in general, take any value in $[0,2\pi)$, it becomes quantised in systems with certain symmetries, such as inversion or chiral symmetry \cite{coutant2024surface, zak1989berry}, leading to robust predictions for edge states in topological insulators.

Beyond its role in topological physics, the Zak phase also naturally connects to the spectral theory of discrete operators. In particular, Weyl $m$-functions arise in the analysis of Jacobi operators and encode detailed spectral information, including the density of states and the location of spectral gaps \cite{Teschl1999-dz,damanik.fillman2022OneDimensional}. This spectral perspective provides a bridge between the abstract mathematical framework and physically observable quantities such as edge modes and transport properties.

In this work, we exploit this connection to provide a novel formula for the Zak phase of discrete tight-binding operators in terms of Weyl $m$-functions, continuing the trend \cite{ammari2025uniformhyperbolicitybandgapsedge,ammari2025topological,huang2025resonanceanalysisonedimensionalacoustic} of rephrasing results from mathematical physics in the language of dynamical systems. This approach not only offers a new computational tool but also sheds light on the dependence of the Zak phase on edge conditions and boundary choices, a feature that is often implicit in traditional formulations.

The paper is organised as follows. In section 2, we introduce the necessary mathematical framework, including the relevant concepts from the spectral theory of Jacobi operators and a brief overview of Weyl $m$-functions. Building on this foundation, section 3 states the main result and provides a brief discussion of its implications, with particular emphasis on the dependence of the Zak phase on edge conditions. In section 4, we demonstrate the consistency of our formulation with the classical Zak phase and illustrate how it naturally leads to the quantisation of the Zak phase for periodic structures whose fundamental cells possess inversion symmetry. Finally, section 5 is devoted to a rigorous proof of the main result.

%% file: preliminaries.tex
\section{Preliminaries}\label{sec: preliminaries}
%\subsection{Jacobi Operators}\label{sec: jacobi}

The main object of interest in this paper is the Jacobi operators $\mathcal{J}: \ell^2(\mathbb{Z})\rightarrow\ell^2(\mathbb{Z})$ defined by
\begin{equation*}
    \mathcal{J}\phi(n) = \bseq{a}{n-1}\phi(n-1) + \bseq{b}{n}\phi(n) + \bseq{a}{n}\phi(n +1),
\end{equation*}
where $\mathbf{a},  \mathbf{b} \in \ell^\infty(\mathbb{Z})$ and $\bseq{a}{n} >0, \bseq{b}{n} \in \mathbb{R}$ for all $n \in \mathbb{Z}$.  

For a thorough introduction,  we refer the reader to \cite{damanik.fillman2022OneDimensional,Teschl1999-dz}. We may also define the semi-infinite Jacobi operators $\mathcal{J}_\pm$ on $\ell^2(\mathbb{N})$ as
\begin{equation} \label{def:j+}
    \mathcal{J}_+\phi(n) = \begin{cases} &\bseq{b}{1}\phi(1) + \bseq{a}{1}\phi(2),\quad n = 1,\\
        &\bseq{a}{n-1}\phi(n-1) + \bseq{b}{n}\phi(n) + \bseq{a}{n}\phi(n+1), \quad n\geq 2,
    \end{cases}
\end{equation}
and
\begin{equation}
    \mathcal{J}_-\phi(n) = \begin{cases} &\bseq{b}{0}\phi(0) + \bseq{a}{-1}\phi(-1),\quad n = 0,\\
        &\bseq{a}{n-1}\phi(n-1) + \bseq{b}{n}\phi(n) + \bseq{a}{n}\phi(n+1), \quad n\leq -1.
    \end{cases}
\end{equation}
Note that $ \mathcal{J}_+$ is defined by \eqref{def:j+} for $n \in \mathbb{N}$ with the convention that $\phi(0)=0$. 

Suppose that $\lambda\in\rho(\mathcal{J})$, where $\rho(\mathcal{J})$ denotes the resolvent set of $\mathcal{J}$. Then there exist two formal solutions $u_\pm(\lambda, \cdot)$ to 
\begin{equation*}
    \mathcal{J}u_\pm = \lambda u_\pm,
\end{equation*}
that are square-summable respectively at $\pm\infty$. The function $u_+$ is often referred to as the stable direction, whereas $u_-$ is the unstable direction. Then we may define the Weyl $m$-functions $m_\pm$ for $\lambda \in \rho(\mathcal{J})$ and $n_0\in \mathbb{Z}$ as
\begin{equation}\label{eq: weyl1}
    m_+(\lambda,n_0) = -\frac{u_+(\lambda,n_0 +1)}{\bseq{a}{n_0}u_+(\lambda,n_0 )} \quad \text{and}\quad m_-(\lambda,n_0) = -\frac{u_-(\lambda,n_0)}{\bseq{a}{n_0}u_-(\lambda,n_0 +1)}.
\end{equation}
Let $e_1 \in \ell^2(\mathbb{N})$ be defined by $e_1(n)= 1$ if $n=1$ and $e_1(n)=0$ otherwise. Let $$\mathcal{J}_{\pm,n_0} = ((S^{*})^{n_0} \mathcal{J} S^{n_0})_\pm$$ with $S$ being the shift operator on $\ell^2(\mathbb{Z})$ given by $S\phi(n)= \phi(n+1)$ and let $S^*$ be the $\ell^2$-adjoint of $S$. 

The Weyl $m$-functions may alternatively be equivalently defined by
\begin{equation}
    m_+(\lambda,n_0) = \langle(\mathcal {J}_{+,n_0}-\lambda)^{-1}e_1,e_1\rangle \quad \text{and}\quad m_-(z,n_0) = \langle(\mathcal {J}_{-,n_0}-\lambda)^{-1}e_1,e_1\rangle
\end{equation}
 with the convention that the scalar product is sesquilinear in its second variable, or by
\begin{equation}
    m_+(\lambda,n_0) = \int_\mathbb{R}\frac{1}{z-\lambda}d\mu_+(n_0,z) \quad \text{and}\quad m_-(z,n_0) = \int_\mathbb{R}\frac{1}{z-\lambda}d\mu_-(n_0,z)
\end{equation}
with $d\mu_\pm(n_0,\cdot)$ being the spectral measure associated with $\mathcal{J_\pm}$ and $n_0$. That is, $m_\pm$ is the Borel transform of the measure $\mu_\pm(n_0,\cdot)$. Consequently,  $m_\pm$  is also a Herglotz function \cite{damanik.fillman2022OneDimensional, Teschl1999-dz}. Hence, the limit
\begin{equation}
    N_\pm(\lambda,n_0) = \lim_{\epsilon\rightarrow+0}m_\pm(\lambda + i\epsilon,n_0)
\end{equation}
exist almost everywhere in $\mathbb{R}$ and $d\mu_{\pm,n_0,acc}(z) = N_\pm(z,n_0)d\mu$.
Weyl $m$-functions also satisfy the following property:
\begin{equation}\label{eq: m_represetation}
    m_+(z,i) = \frac{1}{\bseq{b}{i} - z -\bseq{a}{i+1}^2 m_+(z,i+1)}\quad \text{and} \quad m_-(z,i) = \frac{1}{\bseq{b}{i}-z-\bseq{a}{i-1}^2m_-(z,i-1)},
\end{equation}
which we will use later on. In what follows, we will often omit the index of $m_\pm$ and instead define $m_\pm(\lambda) = m_\pm(\lambda,0).$
% Numerous quantities related to the operators $\mathcal{J}_\pm$ and their spectra can be expressed through $m_\pm$. 

% \textcolor{red}{For instance, it is known from \cite{damanik.fillman2022OneDimensional,Teschl1999-dz} that the density of states can be expressed as follows:
% \begin{equation}
%     N(\lambda) = \frac{1}{\pi}\lim_{\epsilon\rightarrow+0}\Im(m_+(\lambda +i0)).
% \end{equation}}\todo[color=cyan]{this is not the physics density of states we are used to!}

%\subsubsection{Laurent and Toeplitz operators}}

Particular attention will be devoted to Jacobi operators with periodic coefficients. We therefore briefly devote a moment to settle the notation.

\begin{definition}
    Let $\mathbb{X} \in \{\mathbb{N}, \mathbb{Z}\}$ and let $\mathcal{J}_{(+)}:\ell^2(\mathbb{X})\rightarrow\ell^2(\mathbb{X})$ be a Jacobi operator with diagonal and off-diagonal elements $\bseq{b}{i}$ and $\bseq{a}{i}$. We say that $\mathcal{J}$ is $p$-periodic if there exists some $p\in\mathbb{N}$ such that
    \begin{equation}
        \bseq{a}{i+p} = \bseq{a}{i} \quad \bseq{b}{i+p} = \bseq{b}{i},  \quad\forall i\in \mathbb{X.}
    \end{equation}
\end{definition}

The Weyl $m$-functions of periodic Jacobi operators are particularly well behaved. Due to the property \eqref{eq: m_represetation}, they can be represented as solutions to a quadratic equation of the form
\begin{equation}\label{eq: m_quadratic}
    a(\lambda) m_\pm(\lambda)^2 + b(\lambda)m_\pm(\lambda) + c(\lambda) = 0.
\end{equation}
By the implicit function theorem, $m_\pm$ is thus locally analytic, and its poles are a subset of the roots of the coefficient $a(z)$. Therefore, the limit
\begin{equation}
    \lim_{\epsilon\rightarrow0} m_\pm(\lambda + i\epsilon)
\end{equation}
exists for almost every $\lambda \in \mathbb{R}$. In particular, the above implies that the spectral density is also locally analytic. 

Two remarks are now in order.
\begin{remark}
    Alternatively, equation \eqref{eq: m_quadratic} can also be easily derived using the transfer matrix of the system; see, e.g., \cite{damanik.fillman2022OneDimensional}.
\end{remark}
\begin{remark}
    The zeros of the leading coefficient in equation \eqref{eq: m_quadratic} actually yield the Dirichlet eigenvalues of the half-line operator. For periodic operators, it is known that these eigenvalues are either within gaps or at the boundary of the band gaps, which is why the absolutely continuous part of the spectral density is analytic in the interior of bands and has at most $(p-1)$ roots.
\end{remark}

The $p$-periodic Jacobi operators are special in the sense that they can be understood as Laurent or Toeplitz operators, for which there is a vast amount of literature available; see, e.g., \cite{bottcher.silbermann1999Introduction, trefethen.embree2020Spectra,ammari.barandun.ea2024Spectra}. 
\begin{lemma}
    Let $\mathcal{J}$ be a $p$-periodic Jacobi operator. Then there exists some continuous function $f$ from $\mathbb{C}$ into $\mathbb{C}^{p\times p}$, called the symbol of the operators $T(f)$ or $L(f)$ that are respectively the associated Toeplitz and Laurent operators to $\mathcal{J}$, such that $\mathcal{J} = L(f)$ if $\mathbb{X} = \mathbb{Z}$ or $\mathcal{J}_+ = T(f)$ if $\mathbb{X} = \mathbb{N}$. 
\end{lemma}
In particular, it is not difficult to see that the symbol $f$ is equal to the Floquet-Bloch transform of the operator $L(f)$. In the physics literature, the symbol $f$ is thus often referred to as a Bloch Hamiltonian. The formal solutions to $(\mathcal{J}-\lambda)u=0$ can be readily constructed knowing that $(k,v)\in\mathbb{C}\times\mathbb{C}^p$ satisfy 
\begin{equation}
    (f(k)-\lambda)v=0.
\end{equation}
Then we can verify that
\begin{equation}
    u=(\dots,e^{-ik}v,v,e^{ik}v, \dots)^\top
\end{equation}
is a formal solution of $(\mathcal{J}-\lambda)u=0$. Here, the superscript $\top$ denotes the transpose. 

Assuming the operator $\mathcal{J}$ to be self adjoint and $p$-periodic, it is a standard result that there exist $\lambda_1^-\leq\lambda_1^+\leq\dots\leq\lambda_p^-\leq\lambda_p^+\in \mathbb{R}$ such that the spectrum  $\sigma(\mathcal{J})$ of $\mathcal{J}$ is given by 
\begin{equation*}
    \sigma(\mathcal{J}) = \bigcup_i \, (\lambda_i^-,\lambda_i^+).
\end{equation*}
Fix some $1\leq n\leq p$, we say that the $n_0$\textsuperscript{th} band is isolated if 
\begin{equation}
    \lambda_{n-1}^+ < \lambda_{n}^-<\lambda_{n}^+ < \lambda_{n+1}^-.
\end{equation}
From standard perturbation theory, it then follows that the eigenvalue $\lambda_{n}(k)$ in that band is an analytic function of the Bloch momentum $k$ and so are the eigenfunctions $u_{n}(k)$ \cite{kato1995Perturbation, kuchment2016overview}. Thus, we may define the Zak phase \cite{Vanderbilt2018, resta1994macroscopic,zak1989berry} as
\begin{equation}
    \gamma_{n} = \int_0^{2\pi} \mathcal{A}_{n}(k) dk,
\end{equation}
where
\begin{equation}
    \mathcal{A}_{n}(k) = -i\langle u_n(k),\partial_k u_n(k)\rangle
\end{equation}
is the Berry connection. It is well known that the Zak phase is quantised under certain symmetry assumptions such as inversion symmetry.

%\subsection{Zak Phase}
% \todo[color=cyan]{perhaps a bit of an overview?}
% The Zak phase is a topological invariant associated to periodic structures. Assuming that the $n$th band is isolated, one may define it as
% \begin{equation}
%     \gamma_{n} = \int_0^{2\pi} \mathcal{A}_{n}(k) dk,
% \end{equation}
% where
% \begin{equation}
%     \mathcal{A}_{n}(k) = -i\langle u_n(k),\partial_k u_n(k)\rangle_d
% \end{equation}
% is the Berry connection. It is well known that the Zak phase is quantised under certain symmetry assumptions such as inversion symmetry \cite{zak1989berry}.

%Note that we denote the scalar product on $\mathbb{C}^d$ by $\langle \cdot, \cdot \rangle_d$.

%% file: setup_and_main_result.tex
\section{Main Result}
The main result of this paper is the following real space formulation of the 
Zak phase.
\begin{theorem}\label{thm: main}
    Let $\mathcal{J_+}:\ell^2(\mathbb{N})\rightarrow\ell^2(\mathbb{N})$ be a self-adjoint periodic Jacobi operator. Assume that the $n$\textsuperscript{th} band is isolated. 
If we define the Berry connection as
    \begin{equation}
        \mathcal{A}_n(k) = -i\langle u_n(k), \partial_ku_n(k)\rangle,
    \end{equation}
    then it holds that
    \begin{equation}
        \mathcal{A}_n(k) = \frac{\lambda_n'(k)}{2}\frac{\Re(m_+'(\lambda+i0))}{\Im(m_+(\lambda+i0))} - \frac{1}{2}.
    \end{equation}
    In particular, if we denote by $\lambda_{n,\min},\lambda_{n,\max}$ the lower and upper band edges, then the following formula for the Zak phase holds:
    \begin{equation}\label{eq: zak_new}
        \gamma_n = \int_{\lambda_{n,\min}}^{\lambda_{n,\max}} \frac{\Re(m_+'(\lambda_n(k) + i0)}{\Im(m_+(\lambda_n(k)+i0)}d\lambda - \pi.
    \end{equation}
\end{theorem}
We will postpone the proof of Theorem \ref{thm: main} to section \ref{sec: proof} in favour of a brief discussion.

Equation \eqref{eq: zak_new} provides many insights into the nature of the Zak phase, first and foremost it is worth emphasising that equation \eqref{eq: zak_new} is entirely formulated in real space, with the only remaining artifact of the translation invariance being the existence of spectral bands. Though such real space representations already exist in the form of trace formulas \cite{onaya2025formally, Jezequel2022-pv, resta1994macroscopic}, the link to Weyl $m$-functions is new to the best of our knowledge. This is noteworthy, as the Weyl $m$-functions are a spectral quantity, whereas usually the Zak phase is thought to be expressible  in terms of phases.

Moreover, its appearance is consistent with the fact that the Zak phase is not a bulk property \cite{Heeger1988,PhysRevB.89.161117,Atala2013, intercellular,resta1994macroscopic}. Instead, its value depends on the choice of the fundamental cell in the definition of $m_\pm$.

% To remedy this, some effort has been put into splitting the Zak phase into an inter- and intra-cellular part, where the former is independent of the choice of unit cell and the latter captures the classical electronic interpretation of the Zak phase \cite{intercellular}. Without further pursuing this point of view, we note that formula \eqref{eq: zak_new} also admits a natural decomposition into a bulk contribution and an edge contribution. To make this precise, we recall that the spectral measure can be decomposed as 
% \begin{equation}
%     \mu = \mu_{ac} + \mu_{pp},
% \end{equation}
% where
% \begin{equation}
%     d\mu_{acc} = N(\lambda)d\lambda,    
% \end{equation}
% and $\mu_{pp}$ is a finite sum of weighted Dirac deltas. Now write
% \begin{equation}
%     \begin{aligned}
%         m_+'(\lambda) &= m'_{+,acc}(\lambda) + m'_{+,acc}(\lambda)\\
%         &=\int_{\mathbb{R}}\frac{1}{(z-\lambda)^2}d\mu_{acc} + \int_{\mathbb{R}}\frac{1}{(z-\lambda)^2}d\mu_{pp}.
%     \end{aligned}
% \end{equation}
% As the absolutely continuous spectrum is independent of the bulk, so is $m'_{+,acc}$, while $m'_{+,pp}$ captures edge effects.

The Weyl $m$-functions are also related to surface impedance. Surface impedance is commonly defined as the ratio of the field to the flux at some energy (or frequency) $\lambda$. Inside a bandgap, this takes the following form:
\begin{equation}
        Z_R(\lambda) = \frac{u_+(0,\lambda)}{\bseq{a}{0}(u_+(0,\lambda) - u_+(1,\lambda))} \quad \text{and}\quad Z_L(\lambda) = \frac{-u_-(0,\lambda)}{\bseq{a}{0}(u_-(0,\lambda) - u_-(1,\lambda))}.
\end{equation}
We can rewrite the above in terms of Weyl $m$-functions to obtain
\begin{equation}
    Z_R(\lambda) =\frac{1}{\bseq{a}{0} + \bseq{a}{0}^2m_+(\lambda)} \quad \text{and} \quad Z_L(\lambda) =\frac{\bseq{a}{0}m_-(\lambda)}{\bseq{a}{0} + \bseq{a}{0}^2m_-(\lambda)}.
\end{equation}
Since the Weyl $m$-functions are real valued in the bandgaps, only the real parts of the Weyl $m$-functions remain in the above formula, providing a link to the Zak phase through equation \eqref{eq: zak_new}.

%% file: numerics.tex
\section{Three Examples}
\subsection{Su–Schrieffer–Heeger Model}
The Bloch Hamiltonian of the Su–Schrieffer–Heeger (SSH) model takes the form
\begin{align}
    f(k) &=
    \begin{pmatrix}
        0 & t_2 \\
        0 & 0
    \end{pmatrix} e^{-ik}
    +
    \begin{pmatrix}
        0 & t_1 \\
        t_1 & 0
    \end{pmatrix}
    +
    \begin{pmatrix}
        0 & 0 \\
        t_2 & 0
    \end{pmatrix} e^{ik}.
\end{align}
A straightforward diagonalisation shows that the eigenvalues are
\begin{equation}
    \lambda(k) = \pm \lvert t_1 + t_2 e^{-ik}\rvert,
\end{equation}
and we may choose the corresponding normalised eigenvectors as
\begin{equation}
    u_k = \frac{1}{\sqrt{2}}
    \begin{pmatrix}
        -e^{i\phi_k} \\
        1
    \end{pmatrix},
\end{equation}
where the phase $\phi_k$ is defined through the complex argument of
$t_1 + t_2 e^{-ik}$.  Differentiation then gives the Berry connection,
\begin{equation}
    \mathcal{A}(k)
    = -i\langle u(k),\, \partial_k u(k) \rangle
    = -\frac{1}{2}\,\partial_k \phi(k),
\end{equation}
which when integrated over the Brillouin zone leads to the well–known expression
for the Zak phase,
\begin{equation}
    \begin{aligned}
        \gamma &= \frac{1}{2}\bigl(\phi(0) - \phi(2\pi)\bigr) \\
        &= 
        \begin{cases}
            0, & \lvert t_1\rvert < \lvert t_2\rvert, \\
            \pi, & \lvert t_1\rvert > \lvert t_2\rvert.
        \end{cases}
    \end{aligned}
\end{equation}
This yields the familiar distinction between the topologically
trivial and non-trivial phases of the SSH chain.

To compute the Weyl $m$-functions, we first determine the stable direction associated with
$\lambda(k+i\epsilon)$. For this, it suffices to identify a kernel element of
$f(k+i\epsilon)-\lambda(k+i\epsilon)$, since the decaying direction can then be constructed
straightforwardly as in section \ref{sec: preliminaries}. A direct computation shows that the Weyl $m_+$-function is given by
% \begin{equation}
%     u(k+i\epsilon)
%     =
%     \begin{pmatrix}
%         \dfrac{t_1+t_2 e^{-ik+\epsilon}}{\lambda(k+i\epsilon)}\\[0.6em]
%         1
%     \end{pmatrix}
%     \Bigg/
%     \sqrt{
%         \left|
%             \dfrac{t_1+t_2 e^{-ik+\epsilon}}{\lambda(k+i\epsilon)}
%         \right|^2
%         +1
%     }.
% \end{equation}
% It follows that the Weyl--$m$-function is given by
\begin{equation}
    m_+\bigl(\lambda(k+i\epsilon)\bigr)
    =
    -\frac{(t_1+t_2 e^{-ik+\epsilon})\,e^{ik-\epsilon}}
         {t_2\,\lambda(k+i\epsilon)}.
\end{equation}
Taking the limit as $\epsilon\to0$, we obtain
\begin{equation}
    m_+\bigl(\lambda(k)\bigr)
    =
    -\frac{e^{i(\phi(\lambda)+k)}}{t_2}.
\end{equation}
Differentiating this expression with respect to $k$ yields
\begin{equation}
    \frac{\Re\bigl(m_+'(\lambda(k))\bigr)}
         {\Im\bigl(m_+(\lambda(k))\bigr)}
    =
    \phi'(\lambda(k))\lambda'(k)+1.
\end{equation}
Hence, our alternative expression for the Zak phase becomes
\begin{equation}
    \tilde{\gamma}
    =
    \phi(\lambda(\pi))-\phi(\lambda(0)).
\end{equation}
Thus, this formulation is fully consistent with the standard approach and reproduces the
usual Zak phase of the SSH model.

\subsection{Rice-Mele Model}
\begin{figure}
    \centering
    \includegraphics[width=1\linewidth]{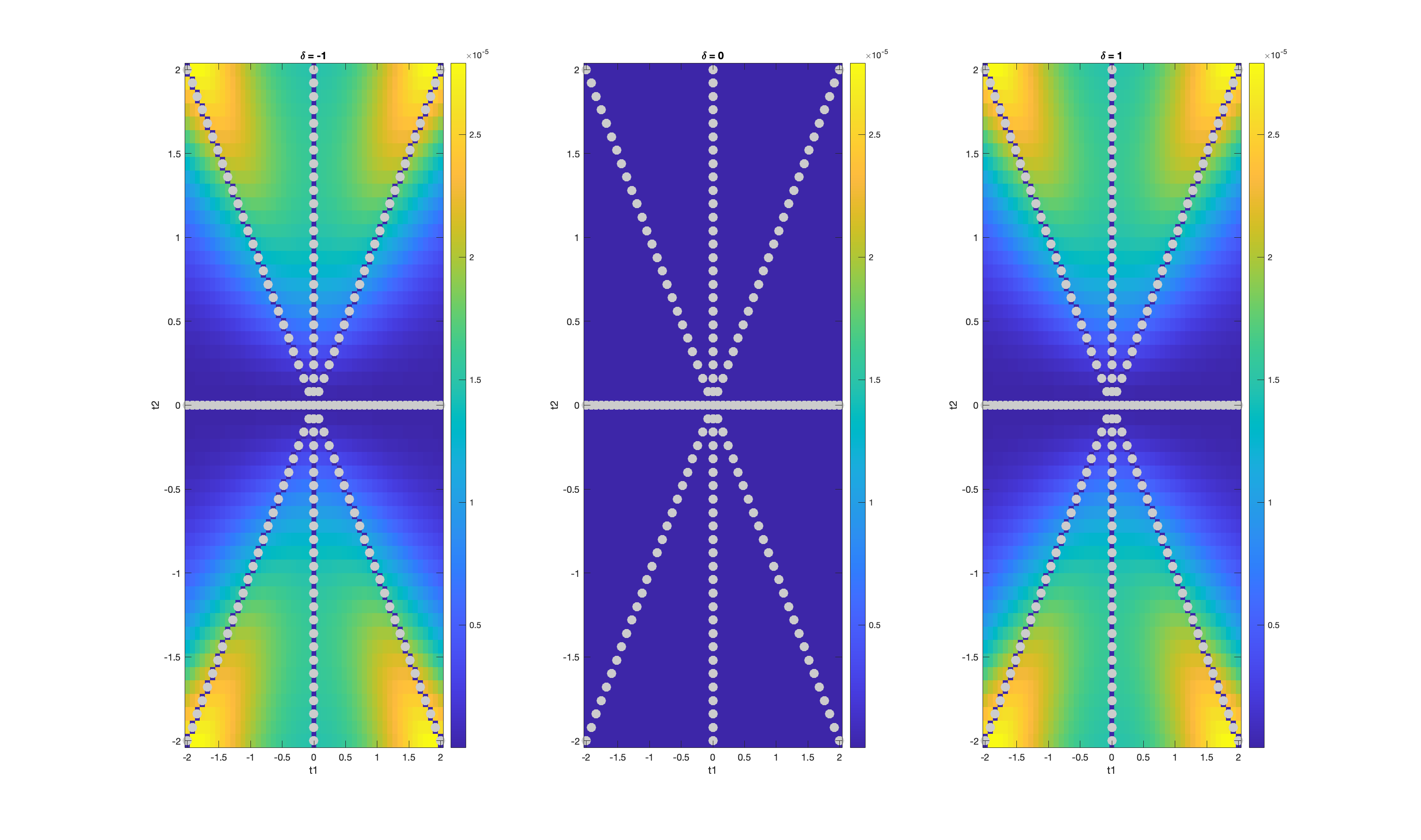}
    \caption{Comparison between formula \eqref{eq: zak_new} and a standard implementation for various parameters. The grey lines mark where the Berry phase becomes degenerate due to equal or vanishing hopping terms. }
    \label{fig: RiceMeleComparison}
\end{figure}

The Rice--Mele model is a one-dimensional tight-binding model that extends the
SSH model by including a staggered on-site potential.
Its Hamiltonian reads
\begin{align}
    H(k) = \begin{pmatrix}
        0& t_2\\
        0 & 0
    \end{pmatrix} e^{-ik}
    + \begin{pmatrix}
        \Delta& t_1\\
        t_1 & -\Delta
    \end{pmatrix}
    + \begin{pmatrix}
        0 & 0\\
        t_2 & 0
    \end{pmatrix} e^{ik},
\end{align}
where \(t_1\) and \(t_2\) denote the intra\-cell and inter\-cell hopping
amplitudes, respectively, and \(\pm\Delta\) is the staggered on-site
potential. In contrast to the SSH model, the presence of a nonzero
\(\Delta\) breaks both the inversion and chiral symmetry. Let $k \in \mathbb{C}_+$ and suppose that $\mathbf{v} =(v_1,v_2)^\top$ solves
\begin{equation}
    H(k)\mathbf{v} = \lambda(k)\mathbf{v}.
\end{equation}
Some algebra then readily yields
\begin{equation}
    \frac{v_1}{v_2} = \frac{\Delta + \lambda}{(t_1 + t_2 e^{ik})}.
\end{equation}
Thus, we obtain an explicit expression for the Weyl-$m_+$ function
\begin{equation}
    m_+(\lambda(k)) = -\frac{\Delta + \lambda}{(t_1 + t_2  e^{ik})}\frac{e^{ik}}{t_2}.
\end{equation}
Thanks to this explicit form, it becomes feasible to accurately evaluate equation \eqref{eq: zak_new} numerically. In Figure \ref{fig: RiceMeleComparison}, we see excellent agreement between the numerical implementation of the new formula and a discrete implementation of the Zak phase as done, e.g., in \cite{Vanderbilt2018}.

\subsection{Mirror Symmetric Fundamental Cells}
It has long been known that inversion symmetric fundamental cells in one-dimensional media lead to a quantisation of the Zak phase \cite{zak1989berry}. This result is usually derived through parity arguments for Wannier functions \cite{zak1989berry} or Bloch eigenfunctions \cite{coutant2024surface}. In this subsection, we show, using only real-space symmetry arguments, how quantisation can be recovered based on Theorem \ref{thm: main}. 
 
\begin{figure}
    \centering
    \includegraphics[width=0.95\linewidth]{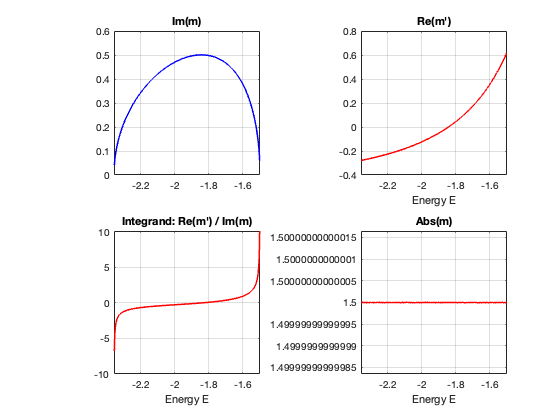}
    \caption{Computation for symmetric trimer setup with $\bseq{b}{1} = \bseq{b}{3}=0, \bseq{b}{2}=1, \bseq{a}{1}=\bseq{a}{2}=1.2$ and $\bseq{a}{3}=1.5$. Using the new formula and $N=500$ discretisation points, the Zak phase of the first band is computed to be $3.1284$, while the standard implementation yields a value of $\pi$.}
    \label{fig: trimer}
\end{figure}

\begin{definition}
    Let $\mathcal{J_+}:\ell^2(\mathbb{N})\rightarrow\ell^2(\mathbb{N})$ be a $p$-periodic Jacobi operator. We say that $\mathcal{J_+}$ has an inversion/mirror symmetric fundamental cell if the following holds:
    \begin{equation}
        \bseq{b}{i} = \bseq{b}{p-i+1} \quad \text{and}\quad\bseq{a}{i} = \bseq{a}{p-i} \quad \text{ for } i=1, \dots, p.
    \end{equation}
\end{definition}
As our goal is to recover the quantisation of the Zak phase using only equation \eqref{eq: zak_new}, we have to understand how symmetry affects the Weyl $m$-functions. This is the purpose of the following lemma. 
\begin{lemma}\label{lem: modulus}
    Let $\mathcal{J_+}:\ell^2(\mathbb{N})\rightarrow\ell^2(\mathbb{N})$ be a $p$-periodic Jacobi operator with an inversion symmetric fundamental cell. Suppose that the $n$\textsuperscript{th} band is isolated. Then $\lvert m_+(\lambda+i0)\rvert = \lvert \bseq{a}{0}\rvert^{-1}$ is constant along that band.
\end{lemma}
The finding of this lemma is highlighted in Figure \ref{fig: trimer}, where, among other things, we see that the absolute value of the Weyl $m$-function is indeed constant for a symmetric trimer setup. In the proof of the above lemma, we will make use of yet another property of periodic Jacobi operators: reflectionless-ness. For our purposes, the following definition of reflectionless is sufficient, though stronger characterisations exist \cite{Teschl1999-dz}.
\begin{definition}
    A Jacobi operator $H$ is called reflectionless if, for all $n\in \mathbb{Z}$ and $\lambda \in \sigma(H)$, the following holds:
    \begin{equation}
        u_+(\lambda + i0,n_0) = \overline{u_-(\lambda + i0,n_0)}.
    \end{equation}
\end{definition}
\begin{proof}
    Without loss of generality, we may assume that $\lambda = 0$. Let $\mathcal{J}: \ell^2(\mathbb{Z}\rightarrow \ell^2(\mathbb{Z})$ be the associated periodic Jacobi operator, where the fundamental cells are arranged so that a fundamental cell begins at $1$. Define the operator $$W:\ell^2(\mathbb{Z})\rightarrow\ell^2(\mathbb{Z}), e_{n}\rightarrow e_{-n+1},$$ where $\{e_n\}_{n\in \mathbb{Z}}$ denotes the standard basis.  If we denote by $u_\pm$ the solution that is in $\ell^2$ as $\pm n\rightarrow \infty$, then we have
    \begin{equation}
        \begin{aligned}
            (\mathcal{J}Wu_+)_n&=\bseq{a}{n-1}u_+(-n+2) + \bseq{b}{n}u_+(-n+1) + \bseq{a}{n}u_+(-n)\\
            &=\bseq{a}{-n+1}u_+(-n+2) + \bseq{b}{-n+1}u_+(-n+1) + \bseq{a}{-n}u_+(-n)\\
            &=0.
        \end{aligned}
    \end{equation}
    Therefore, $Wu$ is also a solution of $\mathcal{J}Wu=0$ and is in $\ell^2$ as $n\rightarrow-\infty$. This implies that $Wu_+$ must be a multiple of $u_-$, that is, $u_+(\lambda,n) = u_-(\lambda,-n+1)$. Because $\mathcal{J}$ is also reflectionless, we have $u_+(\lambda,n_0) = \overline{u_-(\lambda, n_0)}$. Therefore, we have
    \begin{equation}
        \begin{aligned}
            m(\lambda,n_0) &= \frac{u_+(\lambda, n_0+1)}{\bseq{a}{n_0}u_+(\lambda, n_0)}\\
            &=\frac{u_+(\lambda, n_0+1)}{\bseq{a}{n_0}\overline{u_-(\lambda, n_0)}}\\
            &=\frac{u_+(\lambda,n_0+1)}{\bseq{a}{n_0}\overline{u_+(\lambda,n_0+1)}},
        \end{aligned}
    \end{equation}
    from which the claim follows.
\end{proof}

We write the Weyl $m$-function in polar form as
\begin{equation}
    m_+(\lambda)=r(\lambda) e^{i\phi(\lambda)}.
\end{equation}
Substituting this representation into the integrand of our new formula
\eqref{eq: zak_new}, we obtain the following relation:
\begin{equation}
    \frac{\Re(m_+'(\lambda))}{|Im(m_+(\lambda))}=\frac{r'(\lambda)}{r(\lambda)}\cot\phi(\lambda)-\phi'(\lambda).
\end{equation}
For $\lambda\in\sigma(\mathcal J)$, the boundary value
\(r'(\lambda+i0)\) vanishes according to Lemma \ref{lem: modulus}. As a consequence,
the first term does not contribute to the integral, and the expression
for the Zak phase simplifies. We are left with
\begin{equation}
    \gamma
    =-\int_{\lambda_{\min}}^{\lambda_{\max}}\phi'(\lambda)\,\mathrm d\lambda
    =\phi(\lambda_{\min})-\phi(\lambda_{\max}).
\end{equation}
Finally, since the Weyl $m$-function is real-valued in the spectral gaps,
the phases at the band edges satisfy
\begin{equation*}
   \phi(\lambda_{\min}),\ \phi(\lambda_{\max})\in\{0,\pi\}. 
\end{equation*}
It follows immediately that the Zak phase can only take values in integer
multiples of \(\pi\), and is therefore quantised.

%% file: proof_main_result.tex
\section{Proof of Theorem \ref{thm: main}}\label{sec: proof}
The main idea of the proof is that we can use analyticity of the eigenvectors to extend the Berry connection into the complex plane. Once this is done, the proof consists of three main steps: First, we express the Berry connection in terms  of the resolvent of $\mathcal{J}_+$. Second, the resolvent is expressed using spectral integrals to rewrite the expression in terms of $m_+$. Lastly, the limit as one approaches the real axis is taken.
Thus, we begin by defining
\begin{equation*}
    \mathcal{A}_n(k + i\epsilon) = -i\langle u_n(k + i\epsilon),\partial_k u_n(k + i\epsilon)\rangle.
\end{equation*}
As $\lambda(k + i\epsilon)\notin\sigma(\mathcal{J)}$, there exists a solution $u_+(k+i\epsilon)\in\ell^2(\mathbb{Z})$ to $\mathcal{J}u =\lambda(k+i\epsilon)$ that is square-summable as $n\rightarrow+\infty$. From section \ref{sec: preliminaries}, we know that, up to phase and normalisation, $u_+(k + i\epsilon)$ is determined by
\begin{equation} \label{eq: proof0}
    u_+(k+i\epsilon) = (J_+-\lambda(k+i\epsilon))^{-1}e_1.
\end{equation}
Crucially, the first $k$ entries of $u_+$ when defined as in \eqref{eq: proof0} form a solution of 
\begin{equation}\label{eq: floquet_solution}
    f(k+i\epsilon)u(k+i\epsilon) =\lambda(k+i\epsilon)u(k+i\epsilon),
\end{equation}
where $f$ is the symbol associated with $\mathcal{J}$. By understanding $u_+$ as a function of $\lambda$, \eqref{eq: proof0} enforces a periodic phase, after normalising we can use $u_+(k+i\epsilon)$ defined by equation \eqref{eq: proof0} to compute the Berry connection. By viewing $\mathcal{J}_+$ as a Toeplitz operator, it can be shown that $u_+$ takes the form
\begin{equation}\label{eq: proof1}
    u_+ = c \left(u(k+i\epsilon), e^{ik -\epsilon} u(k+i\epsilon), e^{2ik -2\epsilon} u (k+i\epsilon),\dots\right)^\top,
\end{equation}
where $u (k+i\epsilon)$ is a normalised solution of \eqref{eq: floquet_solution} whose phase is in agreement with that of $u_+$ and $c\in \mathbb{R}_+$ is a normalisation constant. We have
\begin{equation}
     \lVert u_+(k+i\epsilon)\rVert^2 = c^2\frac{1}{1- e^{2\epsilon}},
\end{equation}
which yields $$c = \lVert u_+(k+i\epsilon)\rVert \sqrt{1- e^{-2\epsilon}}.$$ Furthermore, we want to express $\partial_k u$ through $\partial_k u_+$. To this end, note, on the one hand, that $\partial_ku_+$ solves
\begin{equation}\label{eq: proof2}
    (\mathcal{J}_+-\lambda(k+i\epsilon))\partial_ku_+(k+i\epsilon) = \lambda'(k + i\epsilon)u_+(k +i\epsilon),
\end{equation}
that is,
\begin{equation}\label{eq: proof3}
    \partial_ku_+ = \lambda'(k+i\epsilon)(\mathcal{J}_+-\lambda(k+i\epsilon))^{-2}e_1.
\end{equation}
On the other hand, we can also use the quasi-periodicity of $u_+$ to gain some more knowledge about the entries of $\partial_k u_+$. Since
\begin{equation}
    u_+(k+i\epsilon,n + k ) = e^{ik-\epsilon} u_+(k+i\epsilon,n),
\end{equation}
we find that
\begin{equation}
    \partial_ku_+(k+i\epsilon,n + k ) = e^{ik-\epsilon} \partial_ku_+(k+i\epsilon,n )+i e^{ik-\epsilon} u_+(k+i\epsilon,n).
\end{equation}
This then yields
\begin{equation}\label{eq: proof4}
    \begin{aligned}
        \partial_ku_+(k+i\epsilon) &= c \left(\left(\partial_k u(k + i\epsilon), e^{ik - \epsilon}  \partial_ku(k+i\epsilon), e^{2ik - 2\epsilon} \partial_ku(k+i\epsilon),\dots\right)^\top\right.\\
        &\left.+ i\left(0, e^{ik-\epsilon} u(k+i\epsilon), 2 e^{2ik-2\epsilon} u(k+i\epsilon) ,\dots\right)^\top\right)\\
        &+c' \left(u(k+i\epsilon), e^{ik -\epsilon} u (k+i\epsilon), e^{2ik -2\epsilon} u (k+i\epsilon),\dots\right)^\top.
    \end{aligned}
\end{equation}
Combining equations \eqref{eq: proof1} and \eqref{eq: proof4}, we obtain
\begin{equation} \label{eq:proofn}
    \begin{aligned}
        \langle u_+(k+i\epsilon),\partial_ku_+(k+i\epsilon)\rangle &= c^2\frac{1}{1-e^{-2\epsilon}}\langle u(k + i\epsilon),\partial_k (u+i\epsilon)\rangle  -ic^2\frac{e^{-2\epsilon}}{(1- e^{-2\epsilon})^2}\\
        &+\frac{cc'}{1- e^{-2\epsilon}},
    \end{aligned}
\end{equation}
where we have used the fact that $u(k+i\epsilon)$ is normalised. Rearranging \eqref{eq:proofn} and using \eqref{eq: proof2} yields
\begin{equation}\label{eq: proof5}
    \begin{aligned}
        \langle u(k + i\epsilon),\partial_k u(kc+i\epsilon)\rangle &= \frac{1}{\lVert u_+(k+i\epsilon)\rVert^2}\langle u_+(k+i\epsilon),\partial_k u_+(k+i\epsilon)\rangle + i\frac{e^{-2\epsilon}}{(1- e^{-2\epsilon})^2}\\
        &- \frac{1}{1- e^{-2\epsilon}}\frac{\partial_k\lVert u_+(k+i\epsilon)\rVert}{\lVert u_+(k+i\epsilon)\rVert}.
    \end{aligned}
\end{equation}
Next, we want to use equations \eqref{eq: proof0} and \eqref{eq: proof3} as well as results from section \ref{sec: preliminaries} to express the Berry connection in terms of spectral integrals. To this end, we first note that
\begin{equation}\label{eq: proof6}
    \begin{aligned}
        \langle u_+(k+i\epsilon), \partial_ku_+(k+i\epsilon)\rangle &= \overline{\lambda'}(k+i\epsilon)\langle(\mathcal{J}_+ - \lambda(k+i\epsilon))^{-1}e_1,(\mathcal{J}_+ - \lambda(k+i\epsilon))^{-2}e_1,\rangle\\
        &=\overline{\lambda'}(k+i\epsilon)\langle (\mathcal{J}_+ - \overline{\lambda}(k+i\epsilon))^{-2}(\mathcal{J}_+ - \lambda(k+i\epsilon))^{-1}e_1,e_1\rangle.
    \end{aligned}
\end{equation}
At this point, we introduce the following integrals which will become useful later on:
\begin{equation}\label{eq: proof7}
    \begin{aligned}
        I_0 (\alpha,\eta) &= \int\frac{(z-\alpha)}{(z-\alpha)^2 + \eta^2}d\mu,\\
        I_1 (\alpha,\eta) 
        &= \int\frac{\eta}{(z-\alpha)^2 + \eta^2}d\mu,\\
        I_2 (\alpha,\eta) &= \int\frac{\eta^3}{((z-\alpha)^2+\eta^2)^2}d\mu,\\
         I_3 (\alpha,\eta)  &= \int\frac{\eta^2(z-\alpha)}{((z-\alpha)^2+\eta^2)^2}d\mu,\\
        I_4 (\alpha,\eta) &= \int \frac{(z-\alpha)^2 - \eta^2}{((z-\alpha)^2+\eta^2)^2}d\mu.
    \end{aligned}
\end{equation}
Here, $d\mu$ is the spectral measure associated with $\mathcal{J}_+$ and $e_1$. Through functional calculus, we can express various quantities related to $u_+$ using these integrals \cite{damanik.fillman2022OneDimensional,Teschl1999-dz}:
\begin{equation}\label{eq: proof8}
    \begin{aligned}
        I_1 (\Re(\lambda),\Im(\lambda)) &= \Im(\lambda)\lVert (\mathcal{J}_+-\lambda)^{-1}e_1\rVert^2 ,\\
        I_2 (\Re(\lambda),\Im(\lambda)) &= -\Im(\lambda)^2\Im\left(\langle (\mathcal{J}_+-\bar{\lambda})^{-2}(\mathcal{J}_+-\lambda)^{-1}e_1, e_1\rangle_+\right) ,\\
        I_3 (\Re(\lambda),\Im(\lambda)) &= \Im(\lambda)^2\Re\left(\langle \mathcal{J}_+-\bar{\lambda})^{-2}(\mathcal{J}_+-\lambda)^{-1}e_1, e_1\rangle_+\right) ,\\
        I_4(\Re(\lambda),\Im(\lambda)) &=\Re(m_+'(\lambda)).
    \end{aligned}
\end{equation}
In particular, it holds that
\begin{equation}\label{eq: proof9}
    \begin{aligned}
        \langle u_+(k+i\epsilon),\partial_k u_+(k+i\epsilon)\rangle &= \overline{\lambda'}(k+i\epsilon)\Im(\lambda(k+i\epsilon))^{-2}\left(I_3(\Re(\lambda(k+i\epsilon)),\Im(\lambda(k+i\epsilon)))\right.\\ 
        &\left.- i I_2(\Re(\lambda(k+i\epsilon)),\Im(\lambda(k+i\epsilon)))\right).
    \end{aligned}
\end{equation}
In the following, we will suppress the arguments of the integrals for better readability. Next, we must take care of $\partial_k\lVert u_+\rVert$. Writing $\alpha = \Re(\lambda(k+i\epsilon))$ and $\eta = \Im(\lambda(k+i\epsilon))$ and using equations \eqref{eq: proof8}, we find that
\begin{equation}\label{eq: proof10}
    \begin{aligned}
        \partial_k\lVert u_+\lVert &= \partial_k\sqrt{\eta^{-1}I_1}\\
        &=\frac{1}{2}\lVert u_+\rVert^{-1}\partial_k(\eta^{-1}I_1)\\\
        &=\frac{1}{2}\lVert u_+\rVert^{-1}\left(-\eta^{-2}\eta'I_1 + \eta^{-1}\partial_kI_1\right).
    \end{aligned}
\end{equation}
Upon exchanging integration and differentiation, we find that
\begin{equation}
    \partial_kI_1 = \eta'I_4+2\eta^{-1}\alpha'I_3.
\end{equation}
Substituting this into \eqref{eq: proof10}, we have
\begin{equation}\label{eq: proof11}
    \begin{aligned}
         \partial_k\lVert u_+\lVert &=\frac{1}{2}\lVert u_+\rVert^{-1}\left(-\eta^{-2}\eta'I_1 + 2\eta^{-2}\alpha'I_3 + \eta'\eta^{-1}I_4)\right)\\
         &=\frac{1}{2}\lVert u_+\rVert\left(-\eta'\eta^{-1} + 2\eta^{-1}\alpha'\frac{I_3}{I_1} + \eta'\frac{I_4}{I_1})\right).
    \end{aligned}
\end{equation}
Combining equations \eqref{eq: proof5}, \eqref{eq: proof9}, and \eqref{eq: proof11} yields
\begin{equation}\label{eq: proof12}
    \begin{aligned}
         \langle u(k + i\epsilon),\partial_k (u+i\epsilon)\rangle &= (\alpha'-i\eta')\eta^{-1}\left(\frac{I_3}{I_1} - i\frac{I_2}{I_1}\right) + i\frac{e^{-2\epsilon}}{(1- e^{-2\epsilon} )^2}\\
         &+\frac{1}{2}\frac{\eta'}{\eta} -\frac{\alpha'}{\eta}\frac{I_3}{I_1} - \frac{\eta'}{2}\frac{I_4}{I_1}\\
         &=i\left(\frac{e^{-2\epsilon}}{(1- e^{-2\epsilon})^2} -\frac{1}{2}\frac{\alpha'}{\eta} + \frac{\alpha'}{2}\frac{I_4}{I_1}- \frac{\eta'}{\eta}\frac{I_3}{I_1}\right),
    \end{aligned}
\end{equation}
where we have used the identity
\begin{equation}
    \frac{I_2}{I_1}=\frac{1}{2}-\frac{\eta}{2}\frac{I_4}{I_1}.
\end{equation}
% \begin{remark}
%     We note that equation \eqref{eq: proof12} already provides a useful consistency check: the right-hand side is purely imaginary, as expected.\todo[color=cyan]{leave this part in?}
% \end{remark}
Assume now that $k \notin\{0,2\pi\}$. Here, $\lambda(\cdot)$ is analytic \cite{kuchment2016overview} and we thus expand 
\begin{equation}
    \lambda(k + i\epsilon) = \lambda(k) + i\epsilon\lambda'(k) + \mathcal{O}(\epsilon^2).
\end{equation}
Combining the above expansion with the fact that $\lambda(k)$ and its derivatives are real together with equation \eqref{eq: proof12}, we obtain the following asymptotics:
\begin{equation}\label{eq: proof13}
    \begin{aligned}
        \langle u(k +i\epsilon),\partial_ku(k + i\epsilon)\rangle_d &= i\left( \frac{1}{2\epsilon}-\frac{1}{2} -  \frac{1}{\epsilon}\left(\frac{1}{2} - \frac{\lambda'(k)\epsilon}{2}\frac{I_4}{I_1}\right) -\frac{\lambda''(k)}{\lambda'(k )}\frac{I_3}{I_1}\right) + \mathcal{O}(\epsilon)\\
        &=i\left(\frac{\lambda'(k)}{2}\frac{I_4}{I_1}-\frac{1}{2}-\frac{\lambda''(k)}{\lambda'(k )}\frac{I_3}{I_1}\right) + \mathcal{O}(\epsilon).
    \end{aligned}
\end{equation}
To obtain the result of Theorem \ref{thm: main}, we need to deal with the term containing $I_3$. First, we note that the measure $\mu$ can be decomposed as
\begin{equation}
    \mu=\mu_{ac} + \mu_{pp},
\end{equation}
where the measure $\mu_{ac}$ has its support contained in the essential spectrum of $\mathcal{J}_+$ and $\mu_{pp}$ is a point measure whose support is constrained to the point spectrum of $\mathcal{J}_+$, which in turn is a subset of the resolvent set of $\mathcal{J}$. In particular, we can show that
\begin{equation}
    d\mu_{ac} = N(\lambda)d\lambda,
\end{equation}
where $N(\cdot)$ is the spectral density associated to $\mathcal{J}_+$.
Let $\delta>0$ be small enough that $[\lambda(k)-\delta,\lambda(k)+\delta]$ lies in the interior of $\sigma(\mathcal{J})$. We have
\begin{equation}\label{eq: proof14}
    \begin{aligned}
        I_3&=\int_\mathbb{R}\frac{\eta^2(z-\alpha)}{((z-\alpha)^2+\eta^2)^2}d\mu\\
        &=\int_\mathbb{R}\frac{\eta^2(z-\alpha)}{((z-\alpha)^2+\eta^2)^2}N(z)dz + \eta^2\sum w_i\frac{(z_i-\alpha)^2-\eta^2}{((z_i-\alpha)^2+\eta^2)^2},
    \end{aligned}
\end{equation}
where $z_i,w_i$ are the enumeration of the point spectra and the associated weight. Because $\alpha = \lambda(k)$ lies in a spectral band, $\lvert z_i-\alpha\rvert>0$ for all $i$. Therefore, the sum in equation \eqref{eq: proof14} clearly vanishes as $\eta\rightarrow0$, which is the case when $\epsilon\rightarrow0$. We now split the remaining integral as follows:
\begin{equation}
    \begin{aligned}
        \int_\mathbb{R}\frac{\eta^2(z-\alpha)}{((z-\alpha)^2+\eta^2)^2}N(z)dz &= \int_{\lambda(k)-\delta}^{\lambda(k)+\delta}\frac{\eta^2(z-\alpha)}{((z-\alpha)^2+\eta^2)^2}N(z)dz\\
        &+ \eta^2\int_{\mathbb{R}\setminus[\lambda(k)-\delta,\lambda(k)+\delta]}\frac{(z-\alpha)}{((z-\alpha)^2+\eta^2)^2}N(z).
    \end{aligned}
\end{equation}
Because $\frac{(z-\alpha)}{((z-\alpha)^2+\eta^2)^2}$ is bounded for $\delta$ small enough and $N(\cdot)$ is integrable, the second integral also tends to 0. Through a change of variables, we now obtain the following identity:
\begin{equation}
    \begin{aligned}
        \int_{\lambda(k)-\delta}^{\lambda(k)+\delta}\frac{\eta^2(z-\alpha)}{((z-\alpha)^2+\eta^2)^2}N(z)dz&=\int_{\eta^{-1}(\lambda(k)-\alpha-\delta)}^{\eta^{-1}(\lambda(k)-\alpha+\delta)}\frac{y}{(y^2+1)^2}N(\eta y + \alpha)dy\\
        &=\int_{\eta^{-1}(\lambda(k)-\alpha-\delta)}^{\eta^{-1}(\lambda(k)-\alpha+\delta)}\frac{y}{(y^2+1)^2}N( \alpha)dy\\
        &+\int_{\eta^{-1}(\lambda(k)-\alpha-\delta)}^{\eta^{-1}(\lambda(k)-\alpha+\delta)}\int_0^y\frac{y}{(y^2+1)^2}N'(\eta x +  \alpha)dxdy.
    \end{aligned}
\end{equation}
As $\epsilon$ tends to 0, the first integral on the right-hand side vanishes due to the symmetry. We proceed with the second integral as follows:
\begin{equation}
    \begin{array}{l}
        \ds \lvert\int_{\eta^{-1}(\lambda(k)-\alpha-\delta)}^{\eta^{-1}(\lambda(k)-\alpha+\delta)}\int_0^y\frac{y}{(y^2+1)^2}N'(\eta x +  \alpha)dxdy\rvert\\
        \nm
        \qquad \ds \leq\int_{\eta^{-1}(\lambda(k)-\alpha-\delta)}^{\eta^{-1}(\lambda(k)-\alpha+\delta)}\lvert\int_0^y\frac{y}{(y^2+1)^2}N'(\eta x +  \alpha) dx\rvert dy\\
        \nm
        \qquad \ds \leq\int_{\eta^{-1}(\lambda(k)-\alpha-\delta)}^{\eta^{-1}(\lambda(k)-\alpha+\delta)}\frac{\eta y^2}{(y^2+1)^2}\sup_{x\in[\lambda(k)-\delta,\lambda(k)+\delta]}\lvert N'(x)\rvert dy.
    \end{array}
\end{equation}
As $N(\cdot)$ is analytic away from the band edges and $\frac{y^2}{(y^2+1)^2}$ is integrable, this integral also vanishes in the limit. Using the following dispersion relation in one dimension \cite{kuchment2016overview}:
\begin{equation}
    \Delta(\lambda(k)) = 2\cos(k),
\end{equation}
we obtain
\begin{equation}
    \lambda'(k) = \frac{\sin(k)}{2\Delta'(\lambda(k))}.
\end{equation}
This implies that $\lambda'$ vanishes only at the band edges. We therefore obtain
\begin{equation}
    \begin{aligned}
        \mathcal{A}(k) &= \lim_{\epsilon\rightarrow0}\mathcal{A}(k+i\epsilon)\\
        &=\frac{\lambda'(k)}{2}\frac{\Re (m_+'(\lambda(k) + i0)}{\operatorname{sign}(\lambda'(k))\Im (m_+(\lambda(k) + i0))}-\frac{1}{2}.
    \end{aligned}
\end{equation}
This proves the first part of Theorem \ref{thm: main}. All that remains is a change of variables:
\begin{equation}
    \begin{aligned}
        \gamma &= \int_{-\pi}^{\pi}\mathcal{A}(k)dk\\
        &=\int_{-\pi}^{\pi}\frac{\lambda'(k)}{2}\frac{\Re (m_+'(\lambda(k) + i0))}{\operatorname{sign}(\lambda'(k))\Im (m_+(\lambda(k) + i0))}-\frac{1}{2}dk\\
        &=\operatorname{sign}(\lambda(\pi)-\lambda(0))\int_0^{\pi}\frac{\lambda'(k)}{2}\frac{\Re (m_+'(\lambda(k) + i0))}{\Im (m_+(\lambda(k) + i0))}dk\\
        &-\operatorname{sign}(\lambda(\pi)-\lambda(0))\int_{-\pi}^{0}\frac{\lambda'(k)}{2}\frac{\Re (m_+'(\lambda(k) + i0))}{\Im (m_+(\lambda(k) + i0))}dk -\pi\\
        &=\operatorname{sign}(\lambda(\pi)-\lambda(0))\int_{\lambda(0)}^{\lambda(\pi)}\frac{1}{2}\frac{\Re (m_+'(\lambda + i0))}{\Im (m_+(\lambda + i0))}d\lambda\\
        &-\operatorname{sign}(\lambda(\pi)-\lambda(0))\int_0^\pi\frac{-\lambda'(-k)}{2}\frac{\Re (m_+'(\lambda(-k) + i0))}{\Im (m_+(\lambda(-k) + i0))}dk-\pi\\
        &=\int_{\lambda_{\min}}^{\lambda_{\max}}\frac{\Re (m_+'(\lambda + i0))}{\Im (m_+(\lambda + i0))}d\lambda-\pi,
    \end{aligned}
\end{equation}
which concludes the proof of Theorem \ref{thm: main}.

%% file: conclusion.tex
\section{Concluding Remarks}
In this work, we established a real-space representation of the Zak phase for periodic Jacobi operators by expressing the Berry connection in terms of the half-line Weyl $m$-function. This leads to an explicit spectral formula for the Zak phase that depends only on boundary values of the resolvent and does not require Floquet-Bloch theory.

To highlight consistency with the classical formulation of Zak phase, we studied the SSH model analytically and the Rice-Mele model numerically. Moreover, we showed that the classical quantisation of structures with inversion symmetric fundamental cells can be recovered from equation \eqref{eq: zak_new} without resorting to momentum-space based arguments.

The present work raises some unanswered questions. Perhaps the most important from a conceptual point of view is the precise interpretation of \eqref{eq: zak_new} as a geometric or topological phase.

% Second, there is the hope of applying \eqref{eq: zak_new} to non-translational invariant systems. If this endeavour proves to be successful, this would have profound consequences for the modern understanding of the Zak phase.